# Consensus Dynamics in a non-deterministic Naming Game with Shared Memory

Reginaldo J. da Silva Filho, Matthias R. Brust and Carlos H.C. Ribeiro
Computer Science Division
Technological Institute of Aeronautics (ITA)
São José dos Campos, Brazil

*Abstract*—In the naming game, individuals or agents exchange pairwise local information in order to communicate about objects in their common environment. The goal of the game is to reach a consensus about naming these objects. Originally used to investigate language formation and self-organizing vocabularies, we extend the classical naming game with a globally shared memory accessible by all agents. This shared memory can be interpreted as an external source of knowledge like a book or an Internet site. The extended naming game models an environment similar to one that can be found in the context of social bookmarking and collaborative tagging sites where users tag sites using appropriate labels, but also mimics an important aspect in the field of human-based image labeling. Although the extended naming game is non-deterministic in its word selection, we show that consensus towards a common vocabulary is reached. More importantly, we show the qualitative and quantitative influence of the external source of information, i.e. the shared memory, on the consensus dynamics between the agents.

*Keywords-distributed collaboration; consensus; naming game; collaborative tagging; shared memory; multi-agent system*

## I. INTRODUCTION

The natural emergence of a common language between individuals still remains an unexplained phenomenon. However, a deeper understanding of the evolutionary processes of language formation is indispensable for developing autonomous multi-agent systems where each agent can potentially have different origins and where no knowledge about the language used in an open-ended environment is provided. To put in a question: How can these agents build a common language through local negations and reach a consensus about the meaning of their vocabulary?

A promising model for a deeper understanding of the common language phenomenon is the *naming game* [1]. It describes a model in which individuals can reach a consensus on how to name different objects. All individuals (or agents) exist in the same environment and sense the same set of objects. The agents are able to invent or create words for the objects. An interaction between two agents is a word transmission from one agent (the *speaker*) to the second agent (the *hearer*). The goal of the game is to reach an agreement between speaker and hearer about the object-word association used for a single object [2]. In this way, a self-organized vocabulary or even a common language with syntactic and semantic levels can be built [3]. The naming game thus is a microscopic model for the interaction dynamics among autonomous agents that communicate without any centralized control [4]. The interesting fact is that a consensus can be reached by means of local interactions. Distributed models can be used to understand, for instance, how large populations reach an agreement with respect to the usage of a certain word, how new language constructs are established, the spreading of rumors and opinions, or even how words propagate in social networks.

Besides its application in modeling the language formation process for individuals, agents or robots (in particular in the field of artificial intelligence), the naming game is of relevance to understand the consensus dynamics of collaborative tagging systems of web sites like *delicious* and *flickr* [5] that have become increasingly popular in recent years. The users of such sites can attach keywords or tags to provide information (e.g., favorite sites on the Internet). In a recent study [6], it is shown how collaborative tagging can lead to both regularities regarding users' activities, tag frequency and keyword usage, and stabilities concerning relative proportions between tags for a given URL and strings that define the location of programs or files in the Internet. Although it is potentially possible to have a constantly increasing number of tags, these findings indicate convergence of a name descriptor (the collection of tags) and the concept (the contents in the location itself).

Different variants of naming games played by humans can help to overcome one of the most challenging problems for search engines: Image labeling. The ESP game [7] aims to use humans' perceptual abilities in order to create valuable output in the process of image labeling. Two players are shown the same image but they are not able to communicate. They are then asked to describe the image with labels under a given time constraint (e.g. Google Image Labeler uses 2 minutes). As soon as they use a common label, it is saved in the database to index the image, the players earn points accordingly and the next image is shown. The objective is to get as many points as possible. While the ESP game is initially designed for a two-player game, in a broader context, the label consensus dynamics of the naming game can be directly used to improve the description accuracy of the images.

Research on the naming game uses mainly the introduced communication model above and focuses on showing its convergence empirically [1,3,8]. The convergence of a deterministic naming game, on the other hand, is mathematically proven in [2]. One common characteristic of these models is that their dynamics are influenced only by the local memories of the agents involved. There is no common access memory, implying that the dynamics of these models is completely uncoupled from any influence of an environment external to the one where the negotiation occurs. The consensus, when reached, is a consensus which belongs to a specific population, and makes sense only in that context. From a sociological point of view, such an arrangement can be plainly artificial, or at least very difficult to establish [9].

The naming game model introduced in this paper differs from these approaches in that it enables agents to access a shared (global) memory with a given probability $p$ (see Fig. 1).

The reason for introducing a shared memory originates from the fact that the real world consists of central access points like books, media, and conferences where individuals build a common vocabulary even without a single interaction. Additionally, often an individual tends to search for an external reference before even emitting a word. The shared memory extension might thus be important for modeling and understanding e.g. the influence of the press and media on the consensus of the group of individuals.

Since classical naming games that allow only local negotiations tend to converge [2], it appears reasonable that an extended version using a shared memory should behave similarly. Although one of our contributions is to show that the extended naming game in fact converges, the focus is on the role and impact of a shared memory on the convergence behavior itself. Knowledge about this influence enables the possibility to control the convergence and, thus, to trigger the outcomes. Another contribution of this paper shows that against our expectations the shared memory is not solely responsible for triggering the consensus word, thus giving importance back to the local interactions.

The paper is organized as follows. A detailed description of the shared memory based model is given in Section II. The model has been implemented as a proof-of-concept prototype. Empirical results are shown and discussed in Section III. Finally, Section IV presents the main conclusions the paper.

## II. A NON-DETERMINISTIC NAMING GAME MODEL

This section describes formally the model proposed in Section I. A population of $N$ agents is considered. Each agent has access to a local memory, which can contain potentially any number of words about a given subject. A word can be a composition of alphabetic elements, but also any other kind of unique identifier. Furthermore, all agents have reading access to the common external (shared) memory. This shared memory contains—prior to the beginning of the game—$C$ distinct words ($C \geq 1$). The objective of the game is to reach a steady state (consensus), i.e. a state in which all agents have the same word in their local memories.

At $t = 0$, all agents have empty local memories. At each successive time step ($t = 1,2,3,...$) two agents are randomly selected, one playing the role of the speaker and the other as the hearer. The negotiation dynamics is as follows:

1. The speaker randomly selects one of the words in its own local memory. If the local memory is empty, two actions are possible depending on probability $p$.
   - With probability $\lambda$ the speaker chooses a word from the shared memory whereby the selected word is added to the local memory of the speaker.
   - With probability $1 - \lambda$ a new word is created locally and selected.
2. The speaker transmits the selected word to the hearer.
3. If the hearer does not have the transmitted word in its local memory, the interaction is considered a failure and the hearer adds the transmitted word to the local memory.
4. If the hearer has the transmitted word in the local memory, two actions are possible depending on the probability $p$. In both cases, the negotiation is considered a success.
   - With probability $\lambda$ the agents involved consult the shared memory. If the transmitted word exists in the shared memory, the speaker and hearer remove all other words from their memories.
   - With probability $1 - \lambda$ both agents remove all words, besides the transmitted one, from their local memories.

The model has three inputs: the number of agents $N$, the probability $\lambda$, and the number of words $C$ in the shared memory. The probability $\lambda$ represents the percentile tendency for the agents to check the shared memory. When the probability $\lambda = 0$, the model is reduced to the standard naming game described e.g. in [10], i.e. without any external influences which are represented by the shared memory. When $\lambda = 1$, the game can be interpreted as a controlled version of the naming game having $C$ possible words.

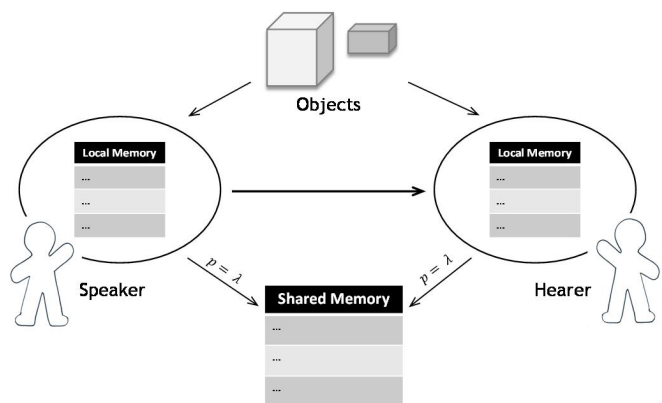

Figure 1. Illustration of the extended naming game using a shared memory that agents are able to access with probability $\lambda$.

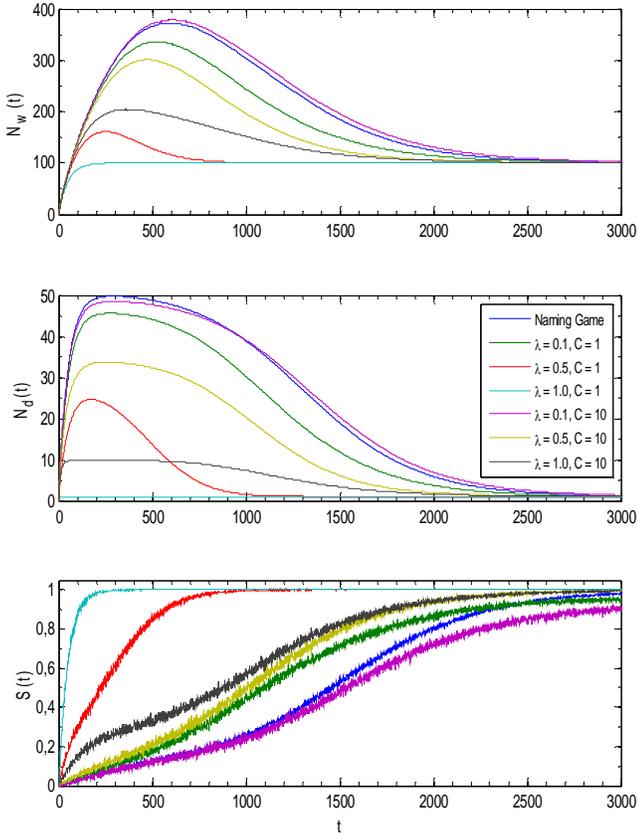

Figure 2. Curves for $N_w$, $N_d$, and $S$ as a function of time for $N = 100$ agents playing the game described by the introduced model.

III. SIMULATION AND RESULTS

The extended naming game communication model has been implemented as a proof-of-concept prototype. Empirical results we obtained by simulation; they are evaluated and discussed in this section.

For the simulation settings, we assume that the agents are in a fully connected network, where each agent can communicate with all others. This is also assumed in the communication models described in [1,2,3,8]. The number of agents $N$ is set to 100 in all simulations. The values of $\lambda$ vary from 0.0 to 1.0, and we tested values of $C$ as 1, 5, 10, 50, 100 and 500. For each combination $(\lambda, C)$ the game was executed 1000 times. The results shown are the averages over these runs.

There are three default measurements for the naming game, see e.g. [11]. The first one is the variation of the number of different words as a function of time $N_w(t)$. For a given time step $t$, the value of $N_w(t)$ is the sum of the number of words in the local memories of all agents. Second, we define $N_d(t)$ as a function that gives the number of different words at time $t$, i.e. it is the number of elements of the set containing all the words in the model at time $t$. Third, we define the success rate $S(t)$ as follows: In a given interaction between two agents, the value 1 is assigned if the interaction is a success and 0 if it is a failure. It is important to note that for a given execution the success rate $S(t)$ can only have values either 0 or 1.

Figure 2 gives an overview of the behavior of the basic properties of the system using the introduced shared memory. It shows that the system dynamics is influenced by the values of both $\lambda$ and $C$. The dark blue curves show the results for the standard naming game [10], which occurs when $\lambda = 0$.

Observing Figure 2, it is possible to see that the system clearly undergoes a disorder/order transition. At the beginning of the game, the total number of words in the system $N_w(t)$ grows smoothly, indicating that unsuccessful interactions occur, a fact that can be confirmed by the low value of $S(t)$. On the other hand, the number of different words $N_d(t)$, grows significantly, quickly reaching its maximum value. This means that new words are introduced. Still at the beginning of the game, the value of $N_d(t)$ begins to decrease, although somewhat moderately, while the value of $N_w(t)$ is still increasing. This means that although successful interactions start to occur, failures are still predominant.

After $N_d(t)$ reaches its maximum value, the word introduction rate slows down. Instead of new word creation, the initially created words spread over the network. The difference of this phase compared to the initial phase is in the fact that the rate for creating new words is steadily decreasing.

The value of $S(t)$ grows moderately at first, but when the existing words are propagated to the majority of the agents, some of them become very popular, and the success rate starts to grow at a faster pace. With the more frequent occurrence of successful interactions, both the total number of words and the number of different words decrease, eventually leading to a consensus state, where $N_d = 1$ and $N_w = N$.

How the input parameters $\lambda$ and $C$ influence the behavior of the system? In other words how the shared memory affects the game dynamics? The most clearly affected property is the maximum value of $N_d(t)$, $max(N_d)$, which is the maximum number of distinct words in the system. For a fixed value of $C$, $max(N_d)$ decreases for increasing values of $\lambda$. On the other hand, for a fixed value of $\lambda$, $max(N_d)$ increases for increasing values of $C$. Figure 3 shows the behavior of $max(N_d)$ with respect to $\lambda$ and $C$.

It is worth noticing that checking of shared memory by the agents as described in Section II is done with probability $\lambda$. The checking can potentially happen in two situations: (a) an agent receives a transmitted word or (b) an agent is selected as speaker and does not have a word in its local memory. In the latter case, the agent can choose one of the words of the shared memory with probability $\lambda$ or invents a new word with probability $1 - \lambda$.

In the classical naming game with $N$ agents [10], $max(N_d)$ is approximately $N/2$. In other words, in the classical naming game on average half of the agents invent new words. This happens because the inventing agents were chosen as speakers while their local memories were empty. With the introduction of the shared memory, this behavior is expected as well, so that on average $N/2$ agents are chosen as speakers while their local memories are empty. Amongst these $N/2$ agents, $\lambda N/2$ choose a word from the shared memory for transmission while $(1 - \lambda) N/2$ will introduce (invent) new words, ideally distinct

ones. Then, the average maximum number of distinct words expected in the system obeys

$$max(N_d(N,\lambda,C)) \leq (1-\lambda)\frac{N}{2} + NC_d(N,\lambda,C),$$

where $NC_d$ represents the maximum possible number of words chosen by the $\lambda N/2$ agents amongst the $C$ words of the shared memory, in other words $NC_d(N,\lambda,C) = C$ if $\lambda N/2 > C$ and $NC_d(N,\lambda,C) = \lambda N/2$, if $\lambda N/2 \leq C$.

Figure 4 shows the variation of the time in which the number of distinct words in the system reaches its maximum value $t_{max(N_d)}$. For a fixed value of $C$, $t_{max(N_d)}$ decreases for increasing values of probability $\lambda$ and for a fixed value of $\lambda$, $t_{max(N_d)}$ increases for increasing values of $C$.

Figure 5 shows the behavior of the average convergence time $t_{conf}$ for the game. We say that the system has converged when every agent has exactly one word, which is the same for all of them, that is, when $N_w = N$ and $N_d = 1$. For $C = 1$ (only one word in the shared memory), the convergence time always decreases when $\lambda$ increases. For other values of $C$, the convergence time is maximum for some $\lambda p$, increasing in the interval $[0, \lambda p)$ and decreasing in $(\lambda p, 1]$. In general, for a fixed value of $\lambda$, $t_{conf}$ increases for increasing values of $C$. When $\lambda = 0$, the convergence time obviously does not depend on $C$ and its value is approximately 2,500, in fact the same registered for the naming game in [10] and, thus, indirectly validating the implementation.

The curves for the maximum number of words in the system, $max(N_w)$ are shown in Figure 6. For a fixed value of $\lambda$, $max(N_w)$ increases for increasing values of $C$. When $C = 1$, the value of $max(N_w)$ always decreases when $\lambda$ increases. For other values of $C$, $max(N_w)$ also reaches its maximum value for some $\lambda p$, increasing in the interval $[0, \lambda p)$ and decreasing in $(\lambda p, 1]$.

Figure 7 shows the behavior of the property $t_{max(N_w)}$, the time in which the total number of words in the system $N_w(t)$ reaches its maximum value. For a fixed value of $\lambda$, $t_{max(N_w)}$ always increases for increasing values of $C$.

For all simulations executed, convergence was observed to a state in which all the agents have the same word, i.e. a steady state. Interestingly the resulting *consensus word* is not always amongst the $C$-words in the shared memory. To analyze this result we define the parameter $P_{shared}$ as the quotient between the number of executions in which the consensus word is also in the shared memory and the total number of executions. It can be understood as the probability that a system with inputs $N$, $\lambda$ and $C$ converges to a word in the shared memory. The behavior of $P_{shared}$ is shown in Fig. 8. Remarkably, the shared memory only contains the consensus word in all executions when $\lambda > 0.5$. For $\lambda < 0.5$ the ratio depends on the number of words $C$ in the shared memory whereby more words mean a lower ratio.

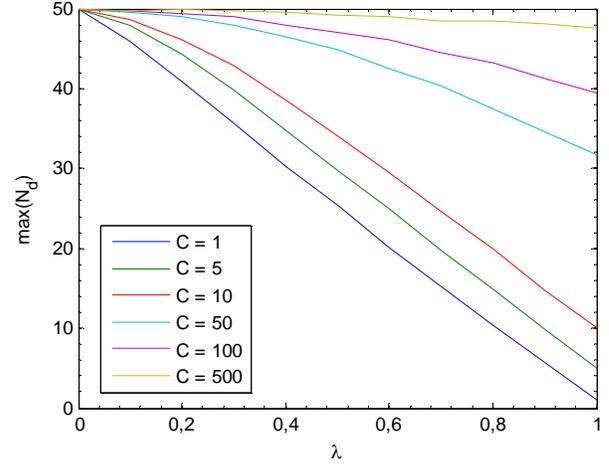

Figure 3. Behavior of the maximum number of distinct words in the system $max(N_d)$ with respect to $\lambda$ and $C$.

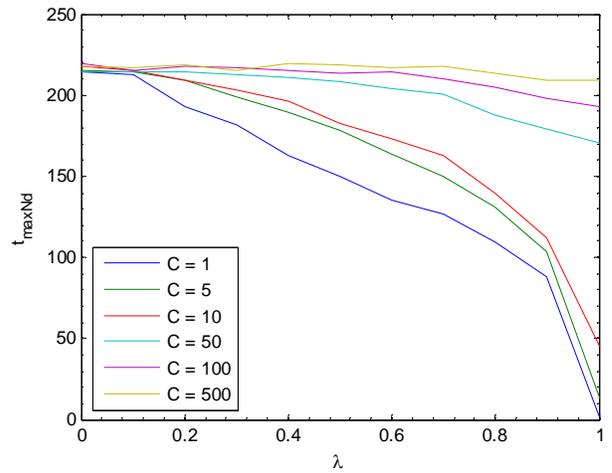

Figure 4. Time when $N_d(t)$ reaches its maximum value $t_{max(N_d)}$.

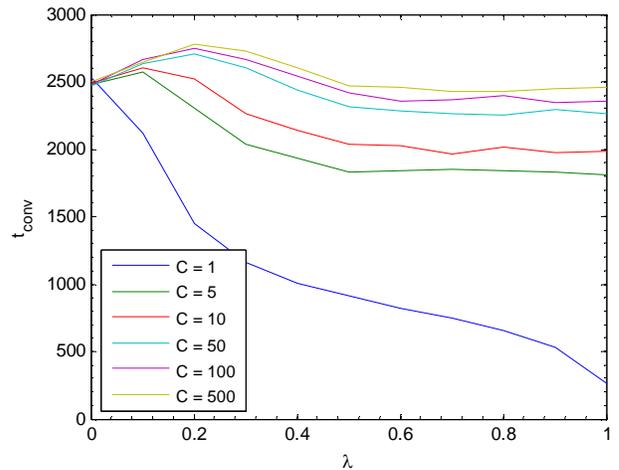

Figure 5. Average convergence times for the proposed model.

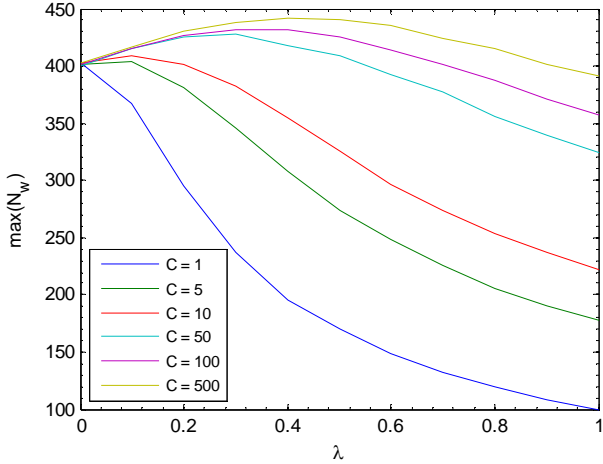

Figure 6. Maximum number of words in the system $max(N_w)$.

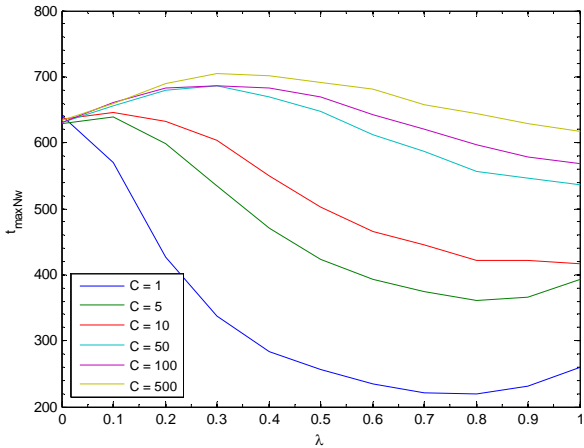

Figure 7. Behavior of the time in which the total number of words in the system is maximum $t_{max(N_w)}$.

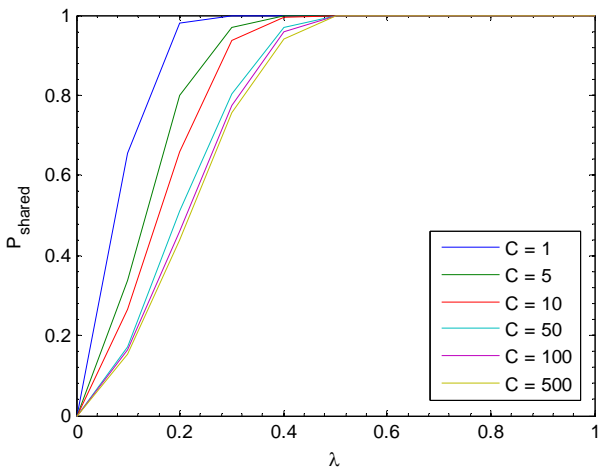

Figure 8. Behavior or property $P_{shared}$ regarding $\lambda$ and $C$.

In order to explain this phenomenon, we consider that the inventions of words only occur in the very beginning of the game. When λ increases, the number of agents which choose words from the shared memory instead of inventing new words increases at the same time. Thus, some of the words which were initially exclusively in the shared memory become very popular from the very beginning of the game. This explains the circumstance that the consensus word has a relatively high probability of belonging to the shared memory.

## IV. DISCUSSION AND CONCLUSION

In this work, we assume that the agents are in a fully connected (mean field) network. However, several works report on how the underlying topology influences the consensus behavior [11,12,13]. Therefore, it is worth investigating how the network topology affects the properties of the extended naming game introduced here. Furthermore, the influence of different selection criteria for the negotiating agents has to be evaluated in order to gain a deeper understanding about the correlation between consensus, determinism, selection probability $\lambda$ and the resulting dynamics.

In the naming game introduced, the agents have only reading access to the shared memory. For further investigations we can consider the idea that some entity has total access to the shared memory that means reading and writing. This entity can determine and manipulate the outcome of the game. Additionally, the value of $\lambda$ does not even have to be the same for all of them. It is acceptable, for instance that each agent $i$ has its own value of $\lambda_i$. Observe that the characteristics of the external shared memory—including but not limited to the value of $C$—are in a certain way determined by the entities that control the external memory. In other words, if these entities allow the agents to access only a limited set of possible words, this results in lowering the value of $C$. On the other hand, with a low value of $C$, it is much easier to predict the outcome of the game. The consensus word will very likely be amongst those words that are interesting to the entities that control the external memory, even if the system has a small value of $\lambda$.

In this paper we introduce a shared external memory into the original naming game communication model [1]. The goal was to analyze how the shared memory can change the dynamics of the naming game regarding the consensus. The external shared memory can be interpreted as the role of a dictionary, a popular reference in a general sense (book, article, encyclopedia, etc.), the news and the press, or a very popular search engine.

The results show that if agents communicate following the rules described in the extended naming game then consensus will always be reached, e.g. the agents reach an agreement on the vocabulary about the objects in their environment. That happens without any centralized control, since the agents only have reading access to the shared memory. We have also empirically shown the degree of impact of the external shared memory on the consensus dynamics. Noteworthy and against expectations the shared memory does not completely determine the consensus word, since simulation showed that the consensus word is not always part of the initial state of the shared memory. This circumstance indicates the importance of local interaction to the common access of a shared memory. In contrast, in this paper we also showed the enormous influence of the shared memory in the consensus dynamics.

ACKNOWLEDGEMENTS

We are grateful to FAPESP and CNPq for supporting the research reported in this paper.

## V. REFERENCES

[1] L. Steels, "The Origins of Ontologies and Communication Conventions in Multi-Agent Systems," *Autonomous Agents and Multi-Agent Systems*, vol. 1, 1998, pp. 169 - 194.

[2] B. De Vylder and K. Tuyls, "How to reach linguistic consensus: a proof of convergence for the naming game.," *Journal of theoretical biology*, vol. 242, 2006, pp. 818-31.

[3] L. Steels, "Self-Organizing Vocabularies," *Proceedings of the Fifth International Workshop on the Synthesis and Simulation of Living Systems*, C.G. Langton and K. Shimohara, Nara, Japan: 1996.

[4] A. Baronchelli, V. Loreto, and L. Steels, "In-depth analysis of the Naming Game dynamics: the homogeneous mixing case," vol. 19, 2008.

[5] C. Marlow, M. Naaman, D. Boyd, and M. Davis, "HT06, tagging paper, taxonomy, Flickr, academic article, to read," *Conference on Hypertext and Hypermedia*, 2006.

[6] S.A. Golder and A. Huberman, "The Structure of Collaborative Tagging Systems," Social Computing Lab, 2005.

[7] L.V. Ahn and L. Dabbish, "Labeling images with a computer game," *Conference on Human Factors in Computing Systems*, 2004.

[8] A. Baronchelli, M. Felici, V. Loreto, E. Caglioti, and L. Steels, "Sharp transition towards shared vocabularies in multi-agent systems," *Journal of Statistical Mechanics: Theory and Experiment*, vol. 2006, 2006, pp. P06014-P06014.

[9] P. Carrington, J. Scott, and S. Wasserman, *Models and Methods in Social Network Analysis*, Cambridge Univ. Press, 2007.

[10] A. Baronchelli, M. Felici, V. Loreto, E. Caglioti, and L. Steels, "Sharp transition towards shared vocabularies in multi-agent systems," vol. 6, 2006.

[11] A. Baronchelli, L. Dall'Asta, A. Barrat, and V. Loreto, "The role of topology on the dynamics of the Naming Game," *The European Physical Journal Special Topics*, vol. 143, 2007, pp. 233-235.

[12] M.R. Brust, C.H. Ribeiro, and J. Mesit, "Avoiding Greediness in Cooperative Peer-to-Peer Networks," *Collaborative Computing: Networking, Applications and Worksharing*, E.B. Joshi and J.B. D., Orlando, USA: Springer Berlin Heidelberg, 2008, pp. 370-378.

[13] A. McIntyre and L. Steels, "Spatially Distributed Naming Games," *Advances in Complex Systems*, vol. 1, 1999, pp. 301-323.